# MCRepair: Multi-Chunk Program Repair
# via Patch Optimization with Buggy Block

Jisung Kim and Byeongjung Lee*

Department of Computer Science and Engineering, University of Seoul
Seoul, Republic of Korea
{kimjisung78, bjlee}@uos.ac.kr

## ABSTRACT

Automated program repair (APR) is a technology that identifies and repairs bugs automatically. However, repairing multi-chunk bugs remains a long-standing and challenging problem because an APR technique must consider dependencies and then reduce the large patch space. In addition, little is known about how to combine individual candidate patches even though multi-chunk bugs require combinations. Therefore, we propose a novel APR technique called multi-code repair (MCRepair), which applies a buggy block, patch optimization, and CodeBERT to target multi-chunk bugs. A buggy block is a novel method that binds buggy chunks into a multi-buggy chunk and preprocesses the chunk with its buggy contexts for patch space reduction and dependency problems. Patch optimization is a novel strategy that effectively combines the generated candidate patches with patch space reduction. In addition, CodeBERT, a BERT for source code datasets, is fine-tuned to address the lack of datasets and out-of-vocabulary problems. We conducted several experiments to evaluate our approach on six project modules of Defects4J. <u>In the experiments using Defects4J, MCRepair repaired 65 bugs, including 21 multi-chunk bugs. Moreover, it fixed 18 unique bugs, including eight multi-chunk bugs, and improved 40–250% performance than the baselines.</u>

**Relationship to the SAC 2023 paper of MCRepair [P]:**

This manuscript reported on the latest experimental results of MCRepair as originally published in the SAC 2023. Because we considered the ACM copyright and could not change the published paper, we only updated the parts related to our experiments with <u>underlines</u> (Revision date: 2023-10-05).

## 4 EXPERIMENT SETUP

### 4.1 Research Questions

MCRepair was implemented for Java. We reported our experiments on the repaired Java bugs in the following sections. Our experiments aimed to answer the following research questions:

**RQ1. What is the performance of MCRepair?**

We evaluated MCRepair on the widely used APR benchmark dataset: Defects4J, to measure its performance.

**RQ2. What is the generalizability of MCRepair?**

We measured generalizability per range of buggy chunks and locations that we used and fixed.

**RQ3. What is the contribution of each component in MCRepair?**

We started with the entire MCRepair technique and removed each component in turn to comprehend its contribution to performance.

### 4.2 Datasets and Ingredients

Bugs2Fix [7] dataset for training and validation and Defects4J [8] dataset for generation and benchmark were used. Especially, Defects4J provides the source codes, testcases, and developer-provided patches. Methods, fields, and their relationships were extracted for ingredients using JavaParser [9] and Spoon [10] libraries. When faults for training and validation, faults were extracted using Java-Diff-Utils [11] library. In addition, faults for generation were extracted using our custom parser.

### 4.3 Fine-tuning and Generation

Considering our hardware specification and the settings of CodeXGlue [12], we set the hyperparameters: embedding_size, max_token_length for a buggy block and its label, training_batch_size, validation_batch_size for fine-tuning, and beam_size for generation are 768, 512, 16, 16, and 500, respectively. The other learning parameters are the same as in [12]. Our learning model was fine-tuned for 100,000 steps using Adam optimizer and $PPL$ (Perplexity) was measured for fine-tuning performance. The lower the value of $PPL$, the better. The minimum $PPL$ value was 1.13909 when the number of training steps reached 100,000. 500 candidate patches were generated per buggy block.

### 4.4 Optimization and Infrastructure

Google-java-format [13] and Java-Parser [9] were used for filtering and ranking, respectively. The action similarity was measured using Gumtree [14] library, where $\alpha$ and $\beta$ were set to 0.5. The n-gram similarity was measured using tri-gram (3-gram). $p$ in Equation 2 was set to 0.5, and $MC$ in Equation 4 was set to 10,000. Our CodeBERT model was implemented using PyTorch [15] and HuggingFace [16]. We evaluated MCRepair on a 16-core server with Ubuntu 18.04 LTS, Docker environment, one NVIDIA RTX A6000 GPU, and two Intel Zeon Gold 6226R 2.9GHz CPUs. We set a time-out per module to 5.5 hours for early termination. It consumed approximately 26 and 807.45 CPU hours for fine-tuning and evaluation, respectively.

# 5 EXPERIMENT RESULTS

The results of MCRepair were opened in the following URL. <u>MCRepair generated 328, 75, and 65 bugs about $CO$, $PL$, and $CR$ results, respectively. In addition, MCRepair built 56.6 K, 1.2 K, and 170 combined patches about $CO$, $PL$, and $CR$ results, respectively.</u>

<u>Recently, we evaluated the patch correctness of MCRepair again. Therefore, we removed two combined patches (i.e., Math-18, Mockito-22) and one bug (i.e., Math-18) that resulted in $CR$ (correctly repaired).</u>

**https://github.com/kimjisung78/MCRepair**

## 5.1 Performance of MCRepair (RQ1)

Most APR studies have focused on the six Defects4J projects [8]: Chart, Closure, Lang, Math, Mockito, and Time. Hence, we used only the project modules with perfect fault localization. Perfect fault localization assumes known buggy locations without fault localization execution. All the results of MCRepair and baselines removed the deprecated modules (i.e., Closure 63, Closure 93, Lang 2, and Time 21) owing to duplication and reproduction problems.

Table 1 shows the comparison between MCRepair and the learning-based and template-based APR baselines with respect to the number of correctly repaired bugs per project module on Defects4J using perfect fault localization. In Table 1, the learning-based baselines are CoCoNut [17], CURE [18], and Recoder [19]. Recoder correctly repaired 71 bugs using perfect fault localization on its paper. However, Recoder opened the revised results in the public [20], and we marked its results as 64 bugs except for the deprecated modules. In Table 1, the template-based baselines are FixMiner [21] and TBar [22]. All template-based baselines were experimented again in [23] using perfect fault localization. Therefore, we marked their results except for the deprecated modules based on [23]. In particular, TBar correctly repaired 74 bugs on its paper, but TBar correctly repaired 53 bugs on [23]. Hence, we marked its results as 52 bugs, excluding the deprecated modules. <u>MCRepair fixed 31, 21, 13, 8, and 1 more bug(s) than FixMiner, CoCoNut, TBar, CURE, and Recoder, respectively.</u>

Table 2 shows the comparison between MCRepair and the state-of-the-art (SOTA) APR baselines with respect to the number of correctly repaired bugs per bug type on Defects4J using perfect fault localization. We classified the results into "<u>Type 1</u>," "<u>Type 2</u>," and "<u>Type 3</u>" using the location-level criterion. The baselines in Table 2 are TBar [22], CoCoNut [17], and CURE [18]. All baselines repaired more than 40 bugs, except for the deprecated modules. In addition, we could classify their results per bug type using the location-level criterion, differ from HERCULES [1], DEAR [24], and Recoder [19]. To clarify the location-level criterion, we did not consider null, blank, and comment locations that were not related to "FAULT_OF_OMISSION," checked whether each location was correct, and distinguished divided and overlapped locations. In addition, a null location is a location that only includes ";" or does nothing (e.g., ";" and "for (int i = 0; i < 10; i++);"). First, "<u>Type 1</u>" is a single-chunk bug that uses or fixes a location. Next, "<u>Type 2</u>" is a single-chunk bug that uses or fixes locations. Finally, "<u>Type 3</u>" is a multi-chunk bug that uses or fixes chunks. The difficulties are in the order of "<u>Type 3</u>," "<u>Type 2</u>," and "<u>Type 1</u>." If an APR technique has "<u>Type 1</u>," "<u>Type 2</u>," and "<u>Type 3</u>" in a module, we resulted in "<u>Type 3</u>," based on the difficulties. For multi-chunk bugs ("<u>Type 3</u>"), CoCoNut, TBar, CURE, and MCRepair fixed 6, 15, 6, and 21 bugs, respectively. MCRepair fixed 6, 15, and 15 more bugs than TBar, CoCoNut, and CURE, (i.e., 40%, 250%, and 250% relative improvements), respectively. In terms of multi-chunk bugs, MCRepair outperformed the baselines. For multi-location bugs ("Types 2-3"), CoCoNut, TBar, CURE, and MCRepair fixed 13, 20, 14, and 28 bugs, respectively. MCRepair fixed 8, 14, and 15 more bugs than TBar, CURE, and CoCoNut, (i.e., 40%, 100%, and 115%), respectively. For total bugs ("Types 1-3"), MCRepair fixed 8, 13, and 21 more bugs than CURE, TBar, and CoCoNut, (i.e., 14%, 25%, and 47%), respectively. In terms of total bugs, MCRepair outperformed the SOTA, learning-based, and template-based baselines. Figure 4 shows the Venn diagram for Table 2. <u>As shown in Fig. 4, MCRepair fixed 18 unique bugs including eight "Type 3" bugs. The results showed that MCRepair was complementary to the SOTA baselines.</u>

***Summary.*** <u>In terms of multi-chunk and multi-location bugs, MCRepair improved 40–250% and 40–115% than the SOTA baselines. Furthermore, in terms of total bugs, MCRepair fixed 65 bugs and outperformed the SOTA, learning-based, and template-based baselines. MCRepair also fixed 18 unique bugs including eight multi-chunk bugs.</u>

## 5.2 Generalizability of MCRepair (RQ2)

All results of RQ2 were the number of bugs that MCRepair correctly fixed, except for the deprecated modules. The statistics of generalizability were based on the buggy locations/chunks that were used and fixed. Figure 5a shows the statistics per chunk range with respect to the number of correctly repaired bugs. <u>MCRepair fixed 44, 18, and 3 bugs about 1, 2, and</u>

**Table 1: RQ1. Comparison with Learning-based and Template-based APR techniques on Defects4J <u>with</u> Perfect Fault Localization.**
**(C: Chart, CL: Closure, L: Lang, M: Math, MC: Mockito, T: Time)**
**The highest number of each column is in bold.**

| Projects | C | CL | L | M | MC | T | Total |
|---|---|---|---|---|---|---|---|
| FixMiner | 7 | 6 | 4 | 12 | 2 | **3** | 34 |
| CoCoNut | 7 | 9 | 7 | 16 | **4** | 1 | 44 |
| TBar | **10** | 13 | **10** | 13 | 3 | **3** | 52 |
| CURE | **10** | 14 | 9 | 19 | **4** | 1 | 57 |
| Recoder | **10** | **21** | **10** | 18 | 2 | **3** | 64 |
| MCRepair | 5 | **21** | 9 | **23** | 4 | **3** | **65** |

**Table 2: RQ1. Comparison with the state-of-the-art APR techniques on Defects4J <u>with</u> Perfect Fault Localization.**
**The highest number of each row is in bold.**

| Bug Types | CoCo-Nut | TBar | CURE | MC-Repair |
|---|---|---|---|---|
| **Type 1. Single-chunk Single-location** | 31 | 32 | **43** | 37 |
| **Type 2. Single-chunk Multi-location** | 7 | 5 | **8** | 7 |
| **Type 3. Multi-chunk Multi-location** | 6 | 15 | 6 | **21** |
| **Total** | 44 | 52 | 57 | **65** |

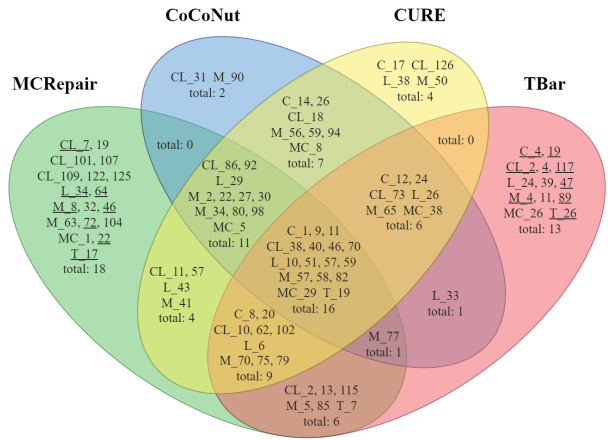

**Figure 4: RQ1. Venn Diagram for Table 2.**
**(Underlined: Unique "Type 3" bugs)**

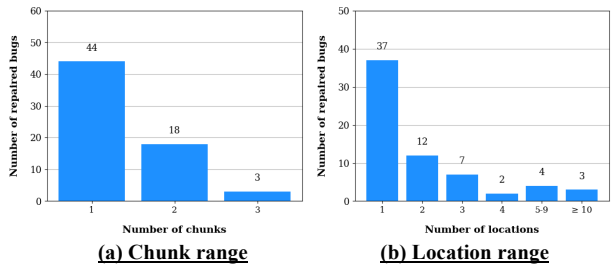

**(a) Chunk range**    **(b) Location range**

**Figure 5: RQ2. Statistics per range on Defects4J with Perfect Fault Localization.**

**TABLE 3: RQ3. Sensitivity Analysis for MCRepair on Defects4J with Perfect Fault Localization.**
**The results are displayed as x/y. x and y are the number of correctly repaired "Type 3" and total bugs.**

| Projects | C | CL | L | M | MC | T | Total |
|---|---|---|---|---|---|---|---|
| −patch optimization | 1/5 | **8**/13 | 3/6 | 1/5 | 0/3 | 1/3 | 18/51 |
| −buggy contexts | 1/6 | **8**/14 | **4**/10 | 6/23 | 0/2 | 1/4 | 20/59 |
| MCRepair | 1/5 | **8**/21 | 3/9 | **7**/23 | 1/4 | 1/3 | **21**/65 |

3 chunks, respectively. In terms of buggy chunks, MCRepair generalized up to three chunks (e.g., Closure-13). Figure 5b shows the statistics per location range with respect to the number of correctly repaired bugs. MCRepair fixed 37, 12, 7, 2, 4, and 3 bugs about 1, 2, 3, 4, 5-9, and 10 or more locations, respectively. In terms of buggy locations, MCRepair generalized up to 13 locations (e.g., Closure-46).

**Summary.** In terms of buggy chunks and locations, MCRepair generalized up to three chunks and 13 locations, respectively.

## 5.3 Contribution of MCRepair's each component (RQ3)

All results of RQ3 were the number of bugs that MCRepair correctly fixed except for the deprecated modules. For the sensitivity analysis, we removed each component from the entire technique one by one. Table 3 lists the contribution of MCRepair's each component with respect to the number of correctly repaired bugs. We removed two components: patch optimization and buggy contexts in buggy blocks. The removal rates decreased by 14 and 6 bugs, respectively. In particular, when MCRepair removed patch optimization, it repaired three "Type 3" fewer bugs. That is the components affected the performance, including multi-chunk bugs.

**Summary.** All components of MCRepair contributed to the performance. Namely, the buggy block and patch optimization that we proposed contributed to the performance including multi-chunk bugs.

## 9 CONCLUSION

In this study, we proposed an APR technique named MCRepair. MCRepair utilized a buggy block, patch optimization, and CodeBERT to target complex multi-chunk bugs. First, a buggy block is a novel method to preprocess buggy chunks into a multi-buggy chunk, considering each dependency to reduce the patch space. Next, patch optimization is a novel strategy to effectively combine candidate patches after filtering and ranking considering the patch space reduction. Finally, we fine-tuned CodeBERT, a BERT model for source code, to supplement the few datasets and OOV problems. In the experiments using Defects4J, MCRepair repaired 65 bugs, including 21 multi-chunk bugs. Moreover, it fixed 18 unique bugs, including eight multi-chunk bugs, and improved 40–250% performance than the baselines. In the future, we plan to solve these limitations (e.g., maximum token length, better relationships, utilization of insertion's ingredients, etc.) and improve MCRepair, considering the findings (e.g., the ability of the applications and the decisions per buggy chunk, the effectiveness of patch space reduction, the testcase terminations, etc.).